\documentclass[12pt,a4paper]{article}%
\usepackage{amsmath,amssymb,amsfonts}
\usepackage{graphicx,epsfig}
\usepackage{hyperref}
\usepackage{color}
\usepackage{amsmath}
\usepackage{amsfonts}
\usepackage{amssymb}
\usepackage{graphicx}
\usepackage{slashed}%
\numberwithin{equation}{section}
\begin{document}
\begin{titlepage}
\begin{flushright}
\end{flushright}
\vskip 1.0cm
\begin{center}
{\Large \bf The Higgs boson from an extended symmetry} \vskip 1.5cm
{\large Riccardo Barbieri, Brando Bellazzini, Vyacheslav S.~Rychkov, Alvise Varagnolo}\\[1cm]
{\it Scuola Normale Superiore and INFN, Piazza dei Cavalieri 7, I-56126 Pisa, Italy} \\[5mm]
\vskip 2.0cm \abstract{The variety of ideas put forward in the
context of a ``composite" picture for the Higgs boson calls for a
simple and effective description of the related phenomenology. Such
a description is given here by means of a ``minimal" model and is
explicitly applied to the example of a Higgs-top sector from an
$SO(5)$ symmetry. We discuss the spectrum, the ElectroWeak Precision
Tests, B-physics and naturalness. We show the difficulty to comply
with the different constraints. The extended gauge sector relative
to the standard $SU(2)\times U(1)$, if  there is any, has little or
no impact on these considerations. We also discuss the relation of
the ``minimal" model with its ``little Higgs" or ``holographic"
extensions based on the same symmetry.}
\end{center}
\end{titlepage}

\section{Introduction}

The first thorough exploration of the energy range well above the Fermi scale,
$G^{-1/2}$, made possible by the Large Hadron Collider, may require a dramatic
revision of the Standard Model (SM) of elementary particles. This is actually
very likely to be the case if the Higgs boson is a naturally light fragment of
the spectrum of whatever theory accounts for the fundamental interactions at
any scale above $G^{-1/2}$. To the point that one wonders whether one should
not have already seen, through the ElectroWeak Precision Tests (EWPT), at
least some indirect manifestation of the required extension of the SM. This
very consideration is in fact at the same time a source of concern and, in
absence of more crucial information, one of the guidelines in trying to
foresee what the LHC will discover.

Without even trying to list the different theoretical directions that have
been taken to address this problem, whose relevance will be judged by the
forthcoming LHC experiments themselves, here we concentrate our attention on
the option that the Higgs boson emerges as a remnant in one way or another of
an (approximate or spontaneously broken) extended symmetry. This is in fact a
rather general framework in itself, with many more specific realizations: the
Higgs as a Pseudo-Goldstone Boson, the little Higgs, the composite Higgs, the
Higgs as $A_{5}$, the intermediate Higgs, the twin Higgs, et cetera. In turn
this variety calls for an effective and, to some extent, unified way to
describe the related relevant phenomenology. Steps in this direction have been
recently made in Ref.~\cite{simplified} and \cite{Rattazzi}. In this work we
focus our attention on the low energy description of the relevant dynamics, as
dictated only by consideration of the approximate symmetry of the Higgs-top
system, since we believe this to be the most important element in judging the
consistency with the current data and in determining the LHC phenomenology. We
look for a simple and, at the same time, accurate description of this dynamics.

A special aspect that emerges from these considerations is the following. In
most of the specific realizations alluded to above, the Higgs boson is thought
to emerge as the low energy remnant of some kind of strong dynamics, hence the
common qualification of \textquotedblleft composite Higgs". While this is
certainly an interesting possibility, actually forced in many specific
realizations by the consistency with the EWPT, we think that it makes also
sense to consider the approximate symmetry of the Higgs-top system as a simple
extension of the SM only, remaining in the perturbative regime. In principle
therefore it is the experiment that should decide between the elementary or
the composite option, leaving open, for the time being, the question of what
will provide the necessary cutoff.

For concreteness we describe in the following a specific example based on the
$SO(5)$ symmetry as a minimal extension of the $SO(4)$ symmetry of the Higgs
potential in the SM. This is obtained by adding to the usual Higgs doublet a
fifth real component and by equally extending the left handed top-bottom
doublet to a five-plet of $SO(5)$. The $SO(5)$ symmetry is allowed to be
broken by soft terms, of unspecified origin, apart from the standard gauge
interactions and the Yukawa couplings other than the top one. The simple
explicit nature of the corresponding Lagrangian allows a straightforward
discussion of the resulting phenomenology, with a special focus on the key
issue, as already mentioned, of the EWPT and of B-physics. Alternative and
relatively more complex ways to extend the top-bottom sector in an
approximately $SO(5)$ invariant way are also described.

Depending on its parameters, our \textquotedblleft minimal" model can either
be viewed as describing a perturbatively coupled Higgs boson or as the low
energy description of a strongly coupled theory at the naturalness cutoff. To
illustrate this dual role we consider the properties of a \textquotedblleft
little Higgs" or a \textquotedblleft holographic" extension. In the last case,
this is precisely the model already discussed in
\cite{Pomarol_spin,Pomarol_fund}, whereas the little Higgs extension
corresponds to the \textquotedblleft deconstructed" version of the same model.

\section{Minimal model at strong coupling}

\label{Strong}Motivated by minimality and by the requirement of including the
custodial symmetry, we consider in the following a model based on the $SO(5)$
symmetry, although the approach followed here is relevant for any model with a
Higgs boson arising from an extended symmetry. As we will explain in more
detail in Section \ref{Non-minimal}, the model gives a low energy description
of any theory in which the ElectroWeak Symmetry Breaking (EWSB) sector has the
$SO(5)$ global symmetry partly gauged by the SM ElectroWeak group. Both for
substantial and for phenomenological reasons we first discuss the
\textquotedblleft strong coupling" case, where the coupling that controls the
$SO(5)\rightarrow SO(4)$ breaking is large. Later we will consider extending
this coupling to the perturbative regime.

\subsection{EWSB sector}

\label{EWSB}The low-energy description of the EWSB sector of our models is the
sigma-model with $SO(5)$ global symmetry broken spontaneously to $SO(4)$. Its
dynamics is described by a scalar five-plet $\phi$ subject to a constraint
\begin{equation}
\phi^{2}=f^{2}, \label{constraint}%
\end{equation}
where $f$ is the scale of the $SO(5)\rightarrow SO(4)$ breaking, which is
assumed to be somewhat higher than the EWSB scale $v=175$ GeV. The cutoff of
this model is%
\begin{equation}
\Lambda\simeq\frac{4\pi f}{\sqrt{N_{g}}}\,, \label{cutoff}%
\end{equation}
where $N_{g}=4$ is the number of Goldstones. One interpretation is that
$\Lambda$ is the compositeness scale of $\phi$, although other UV completions
may be imagined (see Sections \ref{Perturbative},\thinspace\ref{Non-minimal} below).

The SM electroweak group $G_{\text{SM}}=SU(2)_{L}\times U(1)$ gauges a part of
the $SO(5)$. More precisely, we pick a fixed subgroup $SO(4)\equiv
SU(2)_{L}\times SU(2)_{R}\subset SO(5)$, acting on $\vec{\phi}\equiv($the
first 4 components of $\phi$) and gauge $SU(2)_{L}$ and the $T_{3}$ generator
of $SU(2)_{R}$. The kinetic Lagrangian of $\phi$ has thus the form%
\begin{equation}
\mathcal{L}_{\text{kin}}=\frac{1}{2}(D_{\mu}\phi)^{2},\quad D_{\mu}%
\phi=\partial_{\mu}\phi-i(W_{\mu}^{a}T_{L}^{a}+B_{\mu}T_{R}^{3})\phi.
\label{kin}%
\end{equation}

The direction of $\phi$ chooses the angle of alignment between the residual
$SO(4)$ subgroup of the $SO(5)\rightarrow SO(4)$ breaking and the $SO(4)$
inside which $G_{\text{SM}}$ lives. For $\phi=(0,0,0,0,f)$ there is no EWSB
and the W and Z bosons are massless. On the other hand, $\vec{\phi}^{2}=f^{2}$
corresponds to maximal EWSB. In general, we can construct the usual $SU(2)$
Higgs doublet out of $\vec{\phi}$:%
\begin{equation}
H=\frac{1}{\sqrt{2}}\left(
\begin{array}
[c]{c}%
\phi_{1}+i\phi_{2}\\
\phi_{3}+i\phi_{4}%
\end{array}
\right)  . \label{phih}%
\end{equation}
The W boson mass will be related to the Vacuum Expectation Value (VEV) of
$\vec{\phi}$ by the standard relation%
\begin{equation}
m_{W}^{2}=\frac{g^{2}v^{2}}{2},\quad v^{2}=\langle|H|^{2}\rangle=\frac{1}%
{2}\langle\vec{\phi}^{2}\rangle. \label{v1}%
\end{equation}

We will describe the dynamics fixing the VEV\ of $\vec{\phi}$ by a potential
which includes, apart from the $SO(5)$ symmetric term enforcing the constraint
(\ref{constraint}), the most general soft-breaking terms up to dimension 2 and
consistent with the gauge symmetry:%
\begin{equation}
V=V_{0}f^{2}\delta(\phi^{2}-f^{2})-Af^{2}\vec{\phi}^{2}+Bf^{3}\phi_{5}\,.
\label{potential}%
\end{equation}
There may be several sources of these soft-breaking terms (e.g. the gauge
interactions in (\ref{kin}) break the $SO(5)$ symmetry explicitly, and will
generate the $\vec{\phi}^{2}$ term); their precise origin is left unspecified.
We will treat the dimensionless coefficients $A$ and $B$ as free parameters
within their typical ranges consistent with Naturalness as discussed below.

The potential (\ref{potential}) gives a VEV to $\vec{\phi}$ provided that
$A>0$ and
\begin{equation}
\langle\vec{\phi}^{2}\rangle=f^{2}\left[  1-\left(  \frac{B}{2A}\right)
^{2}\right]  >0. \label{vac}%
\end{equation}
This relation shows that to have $v\ll f$ will require finetuning the ratio
$B/2A$ to $1$. This finetuning can be quantifed by the usual logarithmic
derivative\footnote{This estimate will apply e.g.\ under the assumption that
$B$ is distributed uniformly in the range $|B|<2A$. If the typical range of
$B$ is larger, the finetuning will be larger.}:%
\begin{equation}
\Delta=\frac{A}{v^{2}}\frac{\partial v^{2}}{\partial A}\simeq\frac{f^{2}%
}{v^{2}}. \label{ftA}%
\end{equation}
For $f=500$ GeV (the benchmark value used throughout this paper) we have
$\Delta\simeq8$ which corresponds to a $\sim$10\% finetune, and, from
(\ref{cutoff}), to $\Lambda\simeq3~$TeV.

The Higgs particle in this model has a mass%
\begin{equation}
m_{h}=2\sqrt{A}v\text{\thinspace.} \label{mH}%
\end{equation}
Its coupling to the weak gauge bosons, and in fact to any other SM particles,
will be suppressed with respect to the SM by a factor
\begin{equation}
\cos\alpha=\left(  1-\frac{2v^{2}}{f^{2}}\right)  ^{1/2}. \label{cosa}%
\end{equation}
This suppression has its origin in the wavefunction renormalization which
takes place when expanding the kinetic Lagrangian (\ref{kin}) around a point
with $\vec{\phi}\neq0$. Alternatively, it can be viewed as a consequence of
the fact that the Higgs particle is an admixture of an $SU(2)$ doublet
$\vec{\phi}$ and a singlet $\phi_{5}$. At the LHC, such a Higgs boson will
have the VBF (Vector Boson Fusion) production cross section suppressed by
$(\cos\alpha)^{2}\simeq0.75$ for $f=500$ GeV. This effect could be observable,
since the VBF cross section is expected to be measured with $\sim5\%$ error
\cite{Zeppenfeld}.

As a consequence of the reduced coupling of the Higgs particle to the gauge
bosons, the longitudinal $WW$ scattering amplitude grows in this model as
\begin{equation}
\mathcal{A}(W_{L}W_{L}\rightarrow W_{L}W_{L})=-\frac{Gs}{\sqrt{2}}(\sin
{\alpha)}^{2}(1+\cos{\theta})\,, \label{WW}%
\end{equation}
where $s$ is the square of the center-of-mass energy, and $\theta$ is the
scattering angle. This growth can be used to give an alternative estimate for
the cutoff of the model. Indeed, the amplitude (\ref{WW}) would saturate the
unitarity bound at
\begin{equation}
s_{\text{c}}=\frac{s_{\text{c}}^{\text{SM}}}{(\sin{\alpha)}^{2}}\,,
\label{unbound}%
\end{equation}
where $s_{\text{c}}^{\text{SM}}=(1.2~$TeV$)^{2}$ is the analogous bound in the
Higgsless SM\footnote{More precisely, the SM bound is obtained by imposing the
relation $|a_{0}|<1/2$ for the 0th partial wave amplitude for elastic
scattering of the state $(2W_{L}^{+}W_{L}^{-}+Z_{L}Z_{L})/\sqrt{3}$
\cite{elastic}.}. For $f=500$ GeV we have $\sqrt{s_{\text{c}}}=2.4$ TeV, which
is not far from $\Lambda\simeq3$ TeV from (\ref{cutoff}).

\subsection{ElectroWeak Precision Tests}

\label{EWPT}It is straightforward at this point to compute the modifications
introduced in the EWPT relative to the SM, which arise at one loop level due
to the modifed couplings of the Higgs boson to the gauge bosons. Since these
couplings are weaker than standard, the Higgs exchange regulates the
logarithmic divergence in the gauge boson self-energies only partially. The
resulting modification is easy to write down in the heavy Higgs approximation,
in which the electroweak parameters $\hat{S},\hat{T}$ \footnote{We use
parameters $\hat{T},\hat{S}$ \cite{Shat} which are proportional to the
Peskin-Takeuchi parameters: $\hat{T}=\alpha_{\mathrm{EM}}T,\hat{S}=\frac
{g^{2}}{16\pi}S.$} in the SM are given by
\[
\hat{S},\hat{T}=a_{S,T}\log{m_{h}}+b_{S,T}\,,
\]
where $a_{S,T},b_{S,T}$ are constants. In this model, in the same
approximation we will have
\[
\hat{S},\hat{T}=a_{S,T}[(\cos{\alpha})^{2}\log{m_{h}}+(\sin{\alpha})^{2}%
\log{\Lambda}]+b_{S,T}\,,
\]
which amounts to replace $m_{h}$ in the SM by an effective mass
\begin{equation}
m_{\text{EWPT,eff}}=m_{h}(\Lambda/m_{h})^{\sin^{2}{\alpha}}. \label{meff}%
\end{equation}
This modification, numerically important for low $f$, has been typically
overlooked in the previous studies of the composite Higgs models.

On top of this effect we also expect possible contributions from physics at
the cutoff, which can only be estimated by means of proper higher dimensional
operators. We do not expect any such contribution for $\hat{T}$ due to the
custodial $SO(4)$ contained in the $SO(5)$. There will in general be, however,
contributions to $\hat{S}$, which can be estimated as\footnote{This estimate
generally applies in models without T-parity.}
\begin{equation}
\delta\hat{S}|_{\Lambda}\sim\frac{g^{2}v^{2}}{\Lambda^{2}}\simeq
1.4\times10^{-3}\left(  \frac{3~\text{TeV}}{\Lambda}\right)  ^{2}
\label{Scutoff}%
\end{equation}
The study of concrete examples of partial UV completion (see Section
\ref{Non-minimal}) shows that this estimate is trustworthy, including its sign.

\subsection{Naturalness}

\label{Naturalness}

The finetuning estimated in eq. (\ref{ftA}) will be the only source of
finetuning in this model if, as we assume, the parameters $A$ and $B$ take
typical values as consistent with the UV-sensitive contributions from various
couplings breaking the $SO(5)$ symmetry. The parameter $B$ is the only one
which breaks the $\phi_{5}\rightarrow-\phi_{5}$ symmetry\footnote{This
symmetry may be broken by possible Yukawa interactions, see Section
\ref{Third}.} and will be renormalized multiplicatively; there is no
naturalness constraint on its value.

\begin{figure}[h]
\begin{center}
\includegraphics[width=8cm]{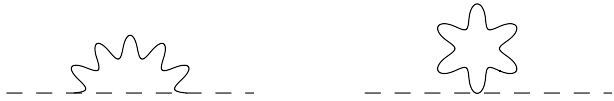}
\end{center}
\caption{{\small Quadratically divergent gauge-boson contributions to the
renormalization of $A$. If the Higgs boson is composite, the scalar form
factor cuts off the divergence in the first diagram. The second diagram is
related to the first one by gauge invariance, and hence will be cut off at a
comparable scale.}}%
\label{gauge-cor}%
\end{figure}

The parameter $A$ is renormalized, first of all, by gauge boson loops:%
\[
\delta A_{\text{gauge}}=-\frac{3(2m_{W}^{2}+m_{Z}^{2})}{16v^{2}}\left(
\frac{\Lambda}{2\pi f}\right)  ^{2}\simeq-0.13\,.
\]
In this estimate we assumed that the loop is cutoff by $\Lambda$, which is
reasonable if the Higgs boson is composite, see Fig.~\ref{gauge-cor}.

Furthermore, if the Yukawa coupling of top is as in the SM, $A$ is also
renormalized by the top quark loop:%
\begin{equation}
\delta A_{\text{top}}=\frac{3m_{t}^{2}}{4v^{2}}\left(  \frac{\Lambda_{top}%
}{2\pi f}\right)  ^{2}\simeq0.7\left(  \frac{\Lambda_{\text{top}}}{3\text{
TeV}}\right)  ^{2} \label{Atop}%
\end{equation}
Although we used $\Lambda_{\text{top}}=3$ TeV as reference, it is worth
pointing out that a priori there is no reason to identify $\Lambda
_{\text{top}}$ with the compositeness scale of the sigma-model $\Lambda$ given
by (\ref{cutoff}). Anyhow, contribution (\ref{Atop}) is the dominant one and
provides a typical expected value of the $A$ parameter.

Via (\ref{mH}), $A\simeq0.7$ corresponds to a 300 GeV Higgs boson. Notice that
we have improved naturalness. Remember that in the SM we have $\Lambda
_{\text{top}}^{\text{SM}}\simeq400$ GeV for $m_{h}=115$ GeV without finetuning
($\Delta=1).$ Here we have increased the Higgs mass and also allowed a
finetuning $\Delta\sim f^{2}/v^{2}$. Thus it is not surprising that we can
raise the scale of physics expected to cutoff the top loop by a factor
$\sqrt{\Delta}(m_{h}/115$ GeV) up to about 3 TeV.

For selfconsistency, we can also estimate the size of the quartic term
$\frac{\kappa}{4}\vec{\phi}^{4}$ omitted from (\ref{potential}). The top loop
will generate a term%
\[
\kappa=\frac{3}{16\pi^{2}}\lambda_{t}^{4}\log\frac{\Lambda^{2}}{m_{t}^{2}%
}\simeq0.1.
\]
Such a small coupling, if present, would be negligible for the present
discussion. In particular, it would not influence in any significant way the
minimization of the potential, and the quadratically divergent contribution to
the $A$ parameter induced by its presence,
\[
\delta A_{\kappa}\approx-\frac{5\kappa}{8}\left(  \frac{\Lambda}{2\pi
f}\right)  ^{2}\simeq-0.06
\]
is negligible compared to $\delta A_{\text{top}}$.

\begin{figure}[h]
\begin{center}
\includegraphics[width=8cm]{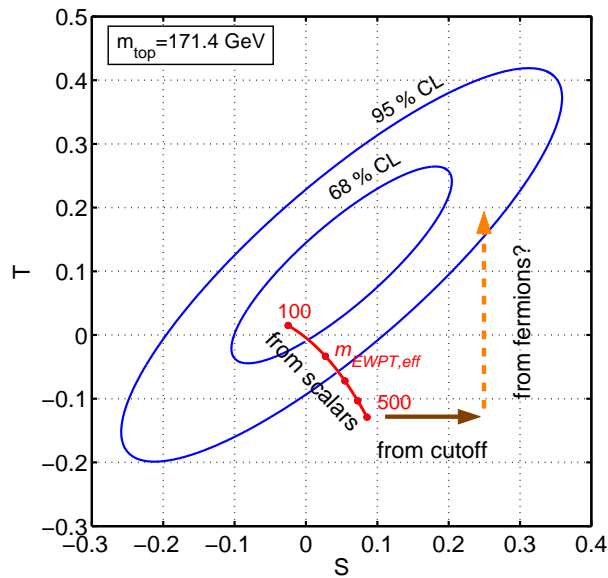}
\end{center}
\caption{{\small The minimal model in the ST plane, including the
contributions (\ref{meff}) (`from scalars') and (\ref{Scutoff}) (`from
cutoff'). The dashed arrow shows an extra positive contribution to $T$ needed
to make the model consistent with the data. In Section \ref{one-loop} we
discuss if such $\delta T>0$ may come from an extended 3rd generation.
Experimental contours taken from the LEPEWWG ST plot \cite{EWWG}. }}%
\label{STplot}%
\end{figure}

As a provisional conclusion, the model as it stands so far is hard to defend
because of the EWPT. For $f=500$ GeV we will have $(\sin\alpha)^{2}%
\simeq0.25,$ $m_{\text{EWPT,eff}}\simeq250\div500$ GeV for $m_{h}=115\div300$
GeV. The combination of (\ref{meff}), (\ref{Scutoff}) leads therefore to an
embarrassing comparison with the experimental constraints on the electroweak
parameters $\hat{S},\hat{T}$. (See Fig.\ \ref{STplot}, the tip of the arrow
marked `from cutoff').

One obvious way to make the model consistent with the EWPT is to increase $f$.
For example, for $f=1$ TeV we will have $(\sin\alpha)^{2}\simeq0.25,$
$m_{\text{EWPT,eff}}\simeq145\div360$ GeV for $m_{h}=115\div300$ GeV, and
$\Delta S\simeq0.04$ from (\ref{Scutoff}). In principle, this is consistent
(at the border of the 2$\sigma$ ellipse) for $m_{h}$ close to the direct lower
bound. However, the finetuning price of $f=1$ TeV from (\ref{ftA}) is
$\sim3\%$, which in our opinion is starting to get uncomfortably
large\footnote{Recall that the MSSM requires $\sim5\%$ finetuning to increase
the Higgs mass above the direct lower bound.}. Because of this we would like
to stick to $f=500$ GeV, and pursue another strategy to improve the EWPT
consistency. Namely, we will add a new sector to the model which provides an
extra positive contribution to $\hat{T}$. In \cite{IDM}, we solved a similar
problem by enlarging the scalar sector of the SM.

\section{Third generation fermions}

\label{Third}

There are two principal motivations for extending the $SO(5)$ symmetry to the
top sector. First of all, the dominant UV sensitivity of the $A$ parameter
given by the top loop, Eq.\ (\ref{Atop}), will be reduced from quadratic to
logarithmic in the extended model. Second, we have seen above that the model
begs for an extra contribution to $\hat{T}$, and it is natural to ask if the
physics which regulates the top loop can be simultaneously responsible for
this contribution. An additional motivation is that we would like to find an
effective four-dimensional (4d) description of existing 5d `composite Higgs'
models \cite{Pomarol_spin}, \cite{Pomarol_fund}, in which the $SO(5)$ symmetry
is naturally present in the fermion sector from the very beginning.

\subsection{ A minimal model}

The minimal way to extend the $SO(5)$ symmetry to the top Yukawa coupling is
to enlarge the left-handed top-bottom doublet $q_{L}$ to a vector $\Psi_{L}$
of $SO(5)$, which under $SU(2)_{L}\times SU(2)_{R}$ breaks up as ($2,2)+1$.
The full fermionic content of the third quark generation will be
\begin{equation}
\Psi_{L}=(q,X,T)_{L};~~~t_{R},b_{R},X_{R},T_{R} \label{f3}%
\end{equation}
where $q_{L},X_{L},X_{R}$ are $SU(2)_{L}$-doublets, while all the other fields
are singlets. We have introduced the right-handed states needed to preserve
parity in the strong and electromagnetic interactions and to give mass to the
new left-handed fermions. The $q_{L}=(t_{L},b_{L})$ and $t_{R},b_{R}$ have the
standard $SU(2)_{L}\times U(1)$ quantum numbers, while the $SO(5)$ symmetry
fixes the hypercharges of the new vector-like states\footnote{The hypercharge
of the components of $\Psi_{L}$ is given by $Y=T_{R}^{3}+2B$, with $B$ the
baryon number. We can take $T_{R}^{3}$ of $q_{L},X_{L}$ as $-1/2$ and $+1/2$
respectively, whereas $T_{L}$ has $T_{R}^{3}=0$.}: $Y(X_{L,R})=7/6$,
$Y(T_{L,R})=2/3.$

The Yukawa Lagrangian of the third quark generation consists of an $SO(5)$
symmetric mass term for the top and of three symmetry-breaking mass terms:
\begin{equation}
\mathcal{L}_{\text{top}}=\lambda_{1}\bar{\Psi}_{L}\phi\,t_{R}+\lambda
_{2}f\,\bar{T}_{L}T_{R}+\lambda_{3}f\bar{T}_{L}t_{R}+m_{X}\,\bar{X}_{L}%
X_{R}+\text{h.c.}\label{Yuk}%
\end{equation}
The coupling $\lambda_{2}$ and the mass $m_{X}$ are soft-breaking terms; at
one loop they generate logarithmically divergent contributions to the $A$
parameter in (\ref{potential}). The coupling $\lambda_{3}$ breaks $\phi
_{5}\rightarrow-\phi_{5}$ symmetry and generates quadratically divergent $B$.
Thus $B\sim A$ can be natural for $\lambda_{3}\sim1/(4\pi)^{2}$. The Yukawa
coupling that generates the bottom mass is taken conventional, i.e. explicitly
breaking, like the gauge couplings, the $SO(5)$ symmetry.

\begin{table}[ptb]%
\begin{equation}%
\begin{array}
[c]{c|c|c}%
\text{state} & \text{mass} & \text{composition}\\\hline
b & 0 & \text{standard}\\[5pt]%
t & m_{t}\simeq\lambda_{t}v(1-\frac{1}{2}\epsilon_{L}^{2}-\frac{1}{2}%
\epsilon_{R}^{2}) &
\begin{array}
[c]{c}%
t_{L}\simeq t_{L}^{0}-\epsilon_{L}T_{L}^{0},\\
t_{R}\simeq t_{R}^{0}-\epsilon_{R}X_{R}^{0}\\[5pt]%
\end{array}
\\
T & m_{T}=\sqrt{\lambda_{1}^{\prime2}+\lambda_{2}^{2}}f & T_{L}\simeq
T_{L}^{0}+\epsilon_{L}t_{L}^{0}\\[5pt]%
X_{5/3} & m_{X} & \text{unchanged}\\[5pt]%
X_{2/3} & m_{X}(1+\frac{1}{2}\epsilon_{R}^{2}) & X_{R}\simeq X_{R}%
^{0}+\epsilon_{R}t_{R}^{0}%
\end{array}
\nonumber
\end{equation}
\caption{{\small The masses and compositions of the physical states in terms
of the fields appearing in (\ref{zero}), denoted by zero superscript. The
mixing parameters are $\epsilon_{R}=\frac{m_{t}}{m_{X}}$ and $\epsilon
_{L}=\frac{\lambda_{T}v}{m_{T}}$ }}%
\label{states}%
\end{table}

Since explicitly
\[
\bar{\Psi}_{L}\phi\,=\bar{q}_{L}H^{c}+\bar{X}_{L}H+\bar{T}_{L}\phi_{5}\,,
\]
after the EWSB $\mathcal{L}_{\text{top}}$ becomes to leading order in $H$
\begin{equation}
\mathcal{L}_{\text{top}}=\lambda_{1}\bar{q}_{L}H^{c}t_{R}+\lambda_{1}\bar
{X}_{L}Ht_{R}+(\lambda_{1}+\lambda_{3})f\bar{T}_{L}t_{R}+\lambda_{2}f\bar
{T}_{L}T_{R}+m_{X}\bar{X}_{L}X_{R}+\text{h.c.} \label{Yuk2}%
\end{equation}
To zeroth order in $v$, mass matrix diagonalization is achieved by the field
rotation:%
\begin{align*}
&  ~T_{R}\rightarrow\cos\chi\,T_{R}-\sin\chi\,t_{R}\,,\quad t_{R}%
\rightarrow\cos\chi\,t_{R}+\sin\chi\,T_{R}\,,\quad\quad\\
&  \qquad\qquad\tan{\chi}=\lambda_{1}^{\prime}/\lambda_{2},\quad\lambda
_{1}^{\prime}=\lambda_{1}+\lambda_{3}\,,
\end{align*}
under which%
\begin{align}
&  \mathcal{L}_{\text{top}}\rightarrow\bar{q}_{L}H^{c}(\lambda_{t}%
t_{R}+\lambda_{T}T_{R})+\bar{X}_{L}H(\lambda_{t}t_{R}+\lambda_{T}T_{R}%
)+m_{T}\bar{T}_{L}T_{R}+m_{X}\bar{X}_{L}X_{R}+\text{h.c.}\label{zero}\\
&  \qquad\qquad\lambda_{t}=\frac{\lambda_{1}\lambda_{2}}{\sqrt{\lambda
_{1}^{\prime2}+\lambda_{2}^{2}}},\quad\lambda_{T}=\frac{\lambda_{1}\lambda
_{1}^{\prime}}{\sqrt{\lambda_{1}^{\prime2}+\lambda_{2}^{2}}},\quad m_{T}%
=\sqrt{\lambda_{1}^{\prime2}+\lambda_{2}^{2}}f\,.\nonumber
\end{align}
The mass and the composition of the physical top quark and of the three new
quarks $T$, $X_{2/3}$ and $X_{5/3}$ in terms of the fields appearing in
(\ref{zero}), in the relevant limit $m_{T,}m_{X}\gg m_{t}$, are given in Table 1.

Notice that the bottom quark in this model is, at tree level, completely
standard, which is just a consequence of the absence of states which could mix
with it.

\subsection{Fermionic loop corrections}

\label{one-loop}The new fermions will give rise to relevant loop corrections,
both to the parameters in the potential (\ref{potential}) and to several
directly observable quantities. Here we concentrate on the contributions to
the electroweak parameters and to B-physics.

\subsubsection{The $\rho$-parameter}

Unlike the scalar sector, in the fermion sector every modification of $\hat
{T},\hat{S}$ relative to the SM dies out as $v^{2}/m^{2}$, where $m$ is a mass
of the new colored states. Nevertheless, since $\hat{T}$, or $\delta\rho$, in
the SM from the top-bottom loops is about $1\%$, the extra correction to
$\hat{T}$ from the heavy fermions may be significant, whereas they are
negligible in the case of $\hat{S}$. For ease of exposition, we give
approximate analytic formulae for the corrections to $\hat{T}$ starting from
(\ref{zero}) with the term $\bar{X}_{L}H\lambda_{T}T_{R}$ neglected. This
allows to treat separately the contributions from $T$ and $X$, which are mixed
with the top via the parameters $\epsilon_{L}=\lambda_{T}v/m_{T}$ and
$\epsilon_{R}=m_{t}/m_{X}$ respectively (See Table \ref{states}). We have
checked numerically that, in the region of interest, these approximations are
defendable and can correctly guide the physical discussion of the various effects.

To leading order in $v^{2}/m^{2}$, the extra contributions to $\hat{T}$ are
given by\footnote{This result can also be found in \cite{Carena}, eq.
(42),(43). We are grateful to Jos\'e Santiago for pointing out a numerical
error in the first version of the paper.}%

\begin{align}
\delta\hat{T}_{T}  &  \simeq\hat{T}_{\text{top}}^{\text{SM}}\left[
2\epsilon_{L}^{2}\left(  \log{\frac{m_{T}^{2}}{m_{t}^{2}}}-1+\frac{\lambda
_{T}^{2}}{2\lambda_{t}^{2}}\right)  \right]  \,,\label{deltaTL}\\
\delta\hat{T}_{X}  &  \simeq\hat{T}_{\text{top}}^{\text{SM}}\left[
-4\epsilon_{R}^{2}\left(  \log{\frac{m_{X}^{2}}{m_{t}^{2}}}-\frac{11}%
{6}\right)  \right]  \,, \label{deltaTR}%
\end{align}
where
\[
\hat{T}_{\text{top}}^{\text{SM}}=\frac{3}{32\pi^{2}}\frac{m_{t}^{2}}{v^{2}%
}\simeq0.009.
\]

Taking into account the starting point of Fig. (\ref{STplot}), the negative
contribution to $\hat{T}$ from (\ref{deltaTR}) makes an $X$-particle lighter
than about $1.5$ TeV unacceptable\footnote{The approximate expressions
(\ref{deltaTL}), (\ref{deltaTR}) are reasonably accurate for $m_{T}\gtrsim500$
GeV, $m_{X}\gtrsim1$ TeV. Numerical analysis shows that $\delta T_{X}$ grows
even more negative for $m_{X}<1$ TeV. (In particular, (\ref{deltaTR}) does not
apply for $m_{X}\lesssim400$ GeV, when the RHS of (\ref{deltaTR}) becomes
positive.) This behavior can be traced to the opposite sign of $T_{3}%
(X_{2/3})$ with respect to the top. Notice that for $m_{X}=0$ Lagrangian
(\ref{Yuk}) is custodially-symmetric, which implies that in this limit the
contribution of $X$ to the $\rho$-parameter should exactly compensate the
standard top contribution.}. On the contrary, a $T$-fermion singlet mixed with
the left handed top can give the desired positive contribution to $\hat{T}$.

\subsubsection{$Z\rightarrow b \bar{b}$ and $b \rightarrow s l \bar{l}$}

\label{Flavor}

Since the top quark mixes with states with different $SU(2)_{L}$ quantum
numbers, see Table \ref{states}, we can expect small deviations from the SM in
effects involving the bottom quark. In the discussion below we will neglect
the mixing between $t_{R}$ and $X_{R}$, since the $X$ quark is necessarily
quite heavy ($m_{X}\gtrsim1.5$ TeV), we do not expect this mixing to lead to
observable constraints.
Instead, we will concentrate on the mixing between $t_{L}$ and $T_{L}$, since
the EWPT suggest that this mixing may be significant.

To have a meaningful unified discussion of all effects, we must first of all
introduce the flavor structure in our model. The most natural way to do this
is to assume that the mechanism described in Section \ref{Third}, with the
softly broken $SO(5)$ symmetric Yukawa coupling for the top, is operational in
the up quark sector of all three generations. For simplicity we also assume
that the matrices $\lambda_{1}$, $\lambda_{2}$ and $\lambda_{3}$ are
simultaneously diagonalizable. On the other hand, the Yukawa couplings
generating the down quark masses are taken like in the SM. Decoupling the very
heavy $X$ quarks, the relevant Yukawa Lagrangian is%
\begin{equation}
\mathcal{L}_{\text{Yuk}}=\bar{q}_{L}H^{c}\lambda^{u}u_{R}+\bar{q}_{L}%
H^{c}\lambda_{T}T_{R}+\bar{T}_{L}m_{T}T_{R}+\bar{q}_{L}HV\lambda^{d}%
d_{R}+\text{h.c.} \label{flav}%
\end{equation}
where we went to the basis in which $\lambda^{u},\lambda^{d},\lambda_{T}%
,m_{T}$ are diagonal, and $V$ is the unitary CKM matrix. The discussion of the
previous subsection remains unchanged with an obvious meaning of the symbols.

As it happens in the SM, one loop exchanges of the top and of the heavier $T$
modify the couplings of the Z to the down quarks as
\[
\left(  -\frac{1}{2}+\frac{\sin^{2}\theta_{\text{w}}}{3}+A_{bb}\right)
\frac{g}{\cos\theta_{\text{w}}}Z_{\mu}\bar{b}_{L}\gamma^{\mu}b_{L}%
\quad\text{and\quad}A_{bs}\frac{g}{\cos\theta_{\text{w}}}Z_{\mu}\bar{b}%
_{L}\gamma^{\mu}s_{L}.
\]
In the large $m_{t}$ limit, the SM values
\[
A_{bb}^{SM}=\frac{\lambda_{t}^{2}}{32\pi^{2}},\quad A_{bs}^{SM}=V_{ts}%
V_{tb}^{\ast}A_{bb}^{SM}%
\]
are corrected in the model under consideration by the same relative factor
\begin{equation}
\frac{A}{A^{SM}}=1+2\epsilon_{L}^{2}\left(  \log{\frac{m_{T}^{2}}{m_{t}^{2}}%
+}\frac{1}{2}+\frac{\lambda_{T}^{2}}{2\lambda_{t}^{2}}\right)  \,.
\label{reduction}%
\end{equation}
The experimental constraints on $A_{bb}$ from the LEP precision measurements
of $R_{b}=\Gamma(Z\rightarrow b\bar{b})/\Gamma(Z\rightarrow$had$)$
is\footnote{We used the measured value $R_{b}=0.21629\pm0.00066$ \cite{Z-pole}
as well as the theoretical estimate $R_{b}=0.21578\pm0.0001-\,0.99(A_{bb}%
-A_{bb}^{SM})$.}%
\begin{equation}
A_{bb}/A_{bb}^{SM}=0.88\pm0.15\,\text{.} \label{Abb}%
\end{equation}
The current constraint on $A_{bs}$ coming from the data on $B\rightarrow
X_{s}l^{+}l^{-}$ decays is \cite{Gino}%
\begin{equation}
A_{bs}/A_{bs}^{SM}=0.95\pm0.20\,\text{.} \label{Abs}%
\end{equation}

A comparison of (\ref{deltaTL}) and (\ref{reduction}) shows that the required
increase of $\hat{T}$ in the SM by about $30-40\%$ will induce an analogous
effect in both $A_{bb}$ and $A_{bs}$ which looks hardly consistent with Eqs
(\ref{Abb}) and (\ref{Abs}). Future measurements of the branching ratio
$B(B_{s}\rightarrow\mu^{+}\mu^{-})$ at the LHCb experiment are expected to
reduce the error in (\ref{Abs}) to 10\% level. A similar effect will also be
present for the $b\to s\gamma$ process, which agrees with the SM at $10\%$
level of error \cite{gambino}.

\subsubsection{$B \bar{B}$ mixing}

In a fully analogous way the box diagrams with exchanges of the top and of the
heavier $T$-quark modify the $B\bar{B}$ mixing both in the $B_{d}$ and in the
$B_{s}$ systems. Still neglecting corrections vanishing like $(m_{W}%
/m_{t})^{2}$, the effective $\Delta B=2$ Lagrangian will have the form, at the
top mass scale,
\begin{equation}
\mathcal{L}^{eff}=C_{d}(\bar{b}_{L}\gamma^{\mu}d_{L})^{2}+C_{s}(\bar{b}%
_{L}\gamma^{\mu}s_{L})^{2} \label{BB}%
\end{equation}
with
\begin{equation}
C_{d,s}=C_{d,s}^{SM}\left[  1+2\epsilon_{L}^{2}\left(  \log{\frac{m_{T}^{2}%
}{m_{t}^{2}}+1}+\frac{\lambda_{T}^{2}}{2\lambda_{t}^{2}}\right)  \right]
\label{BBbar}%
\end{equation}
in terms of the SM coefficients $C_{d,s}^{SM}$. Again, if one wants to produce
the desired $\hat{T}$ from $T$-quark exchanges, an increase in $C_{d,s}$ is
required which may be however at the level of the current $25-30\%$
uncertainty of the lattice calculations of the relevant matrix elements
\cite{UT}.

\section{Third generation quarks: alternatives}

\label{Alternatives}


As we have seen in the previous section, in the model with the minimal fermion
content getting a positive contribution to $T$ of the necessary size seems
impossible without generating at the same time contributions to B-physics
observables exceeding the experimental constraints. One may wonder if this
problem could be solved at the price of extending the 3rd generation sector
even further. Below we will discuss two possible extensions, which have a
common feature that the Yukawa Lagrangian of the 3rd generation consists of
two parts%
\begin{equation}
\mathcal{L}_{\text{top}}=\mathcal{L}_{\text{int}}+\mathcal{L}_{\text{BSM}}
\label{Lcust}%
\end{equation}
where the $\mathcal{L}_{\text{BSM}}$ involves only the new,
Beyond-Standard-Model, fermionic fields, while $\mathcal{L}_{\text{int}}$
couples bilinearly the SM fermions to a subset of BSM fields, $Q_{R}, T_{L}$,
with appropriate quantum numbers:
\[
\mathcal{L}_{\text{int}}=\lambda_{1}f\bar{q}_{L}Q_{R}+\lambda_{2}f\bar{T}%
_{L}t_{R}+\text{h.c.}%
\]

\subsection{Fermions in the spinorial}

An alternative description of the top Yukawa coupling in an $SO(5)$-symmetric
way is through the spinor, $\chi$, rather than the vector, $\Psi$, of $SO(5)$
(see e.g. \cite{Intermediate} for a related model). In this case the full
fermionic content in (\ref{f3}) gets replaced by
\[
\chi_{L,R}=(Q,B,T)_{L,R};~~~q_{L},t_{R},b_{R}%
\]
where, under $SU(2)_{L}\times U(1)$,
\[
Q_{L},Q_{R},q_{L}=2_{1/6},~~~B_{L},B_{R},b_{R}=1_{-1/3},~~~T_{L},T_{R}%
,t_{R}=1_{2/3}.
\]
The Yukawa Lagrangian has the form (\ref{Lcust}) with%
\begin{equation}
\mathcal{L}_{\text{BSM}}=y\bar{\chi}_{L}\phi_{M}\Gamma_{M}\,\chi_{R}%
+m_{Q}\,\bar{Q}_{L}Q_{R}+m_{T}\,\bar{T}_{L}T_{R}+m_{B}\,\bar{B}_{L}%
B_{R}+\text{h.c.,} \label{Yuks}%
\end{equation}
where, in terms of the $SO(5)$ $\Gamma$-matrices,
\[
\bar{\chi}_{L}\phi_{M}\Gamma_{M}\,\chi_{R}=f(\bar{Q}_{L}Q_{R}-\bar{B}_{L}%
B_{R}-\bar{T}_{L}T_{R})+\sqrt{2}(\bar{Q}_{L}H^{c}T_{R}+\bar{Q}_{L}HB_{R}%
-\bar{T}_{L}HQ_{R}-\bar{B}_{L}H^{c}Q_{R}).
\]
From these equations it is straightforward to obtain the spectrum and the
composition of all the colored states (4 more than normal). Without doing this
here explicitly, we limit ourselves to notice that the physical left-handed
b-quark becomes an admixture of doublet and singlet $B_{L}$, at first order in
$v/f$, which is phenomenologically problematic for low $f$.


\subsection{Extended model with fermions in the fundamental}

The 3rd quark generation in such a model includes, apart from the SM fields
$q_{L},t_{R},b_{R}$, 5 new quarks organized in a Dirac fiveplet $\Psi
_{L,R}=(Q,X,T)_{L,R}$ of the $SO(5)$ symmetry. The Yukawa Lagrangian has the
form (\ref{Lcust}) with%
\[
\mathcal{L}_{\text{BSM}}=y_{1}\bar{\Psi}_{L}\phi\,T_{R}+y_{2}\bar{T}_{L}%
\phi^{\dagger}\Psi_{R}+m_{Q}\,\bar{Q}_{L}Q_{R}+m_{X}\,\bar{X}_{L}X_{R}%
+m_{T}\bar{T}_{L}T_{R}+\text{h.c}.
\]
Since $b_{L}$ in this model can mix only with $Q_{L}^{d}$, which has the same
quantum numbers, the Z-boson coupling of the physical left-handed bottom quark
will be standard at tree level\footnote{Incidentally, for $m_{Q}=m_{X}$ the
BSM sector of the model has $O(4)=SU(2)_{L}\times SU(2)_{R}\times P_{LR}$
symmetry, which is known to protect the $Zb_{L}b_{L}$ coupling \cite{Zbb-cust}%
.}.

It would be interesting to compute the one-loop contributions to the $T$
parameter in this model explicitly, and determine if there are regions of the
parameter space consistent with the EWPT and the B-physics constraints. We do
not expect, however, that the situation will be significantly better than for
the minimal fermionic content. The basic reason is that in both cases the
major source of positive $\delta T$ is the mixing of $t_{L}$ with the singlet
$T_{L}$, and it is precisely this mixing which, at one-loop level, led to
unacceptably large contributions to $Z\rightarrow b\bar{b}$ and $B\bar{B}$
mixing observables in Section \ref{one-loop}.

\section{Perturbative minimal model}

\label{Perturbative}As already mentioned, a natural interpretation of the
model described so far is in terms of a \textquotedblleft composite" picture
for the Higgs boson, produced by an unspecified strong dynamics at $\Lambda$.
As an alternative, however, it makes sense to consider also the case in which
the entire model is fully perturbative up to a suitable cut-off scale. To this
end we replace the potential (\ref{potential}) with
\begin{equation}
V=\lambda(\phi^{2}-f^{2})^{2}-Af^{2}\vec{\phi}^{2}+Bf^{3}\phi_{5}, \label{pot}%
\end{equation}
where the coupling $\lambda$ is somewhat greater than $A$ and $B$ but always
perturbative. The explicit discussion of this case proceeds along parallel
lines to the ones followed so far for the strong coupling, with one more
parameter present, which is usefully taken as $A/\lambda$. By requiring that
the one loop correction to the squared mass of ${\phi}$ from the symmetric
coupling $\lambda$ does not exceed the tree level term (a weaker condition can
be easily implemented), one obtains the new cutoff scale
\begin{equation}
\Lambda_{\text{nat}}\simeq\frac{4\pi f}{\sqrt{N+2}}\simeq4.7f\,, \label{Lnat}%
\end{equation}
where $N=5$, i.e. $\Lambda_{\text{nat}}\simeq2.4~$TeV for $f=500~$GeV. This
constraint on the cutoff dominates over every other consideration.

\begin{figure}[h]
\begin{center}
\includegraphics[width=8cm]{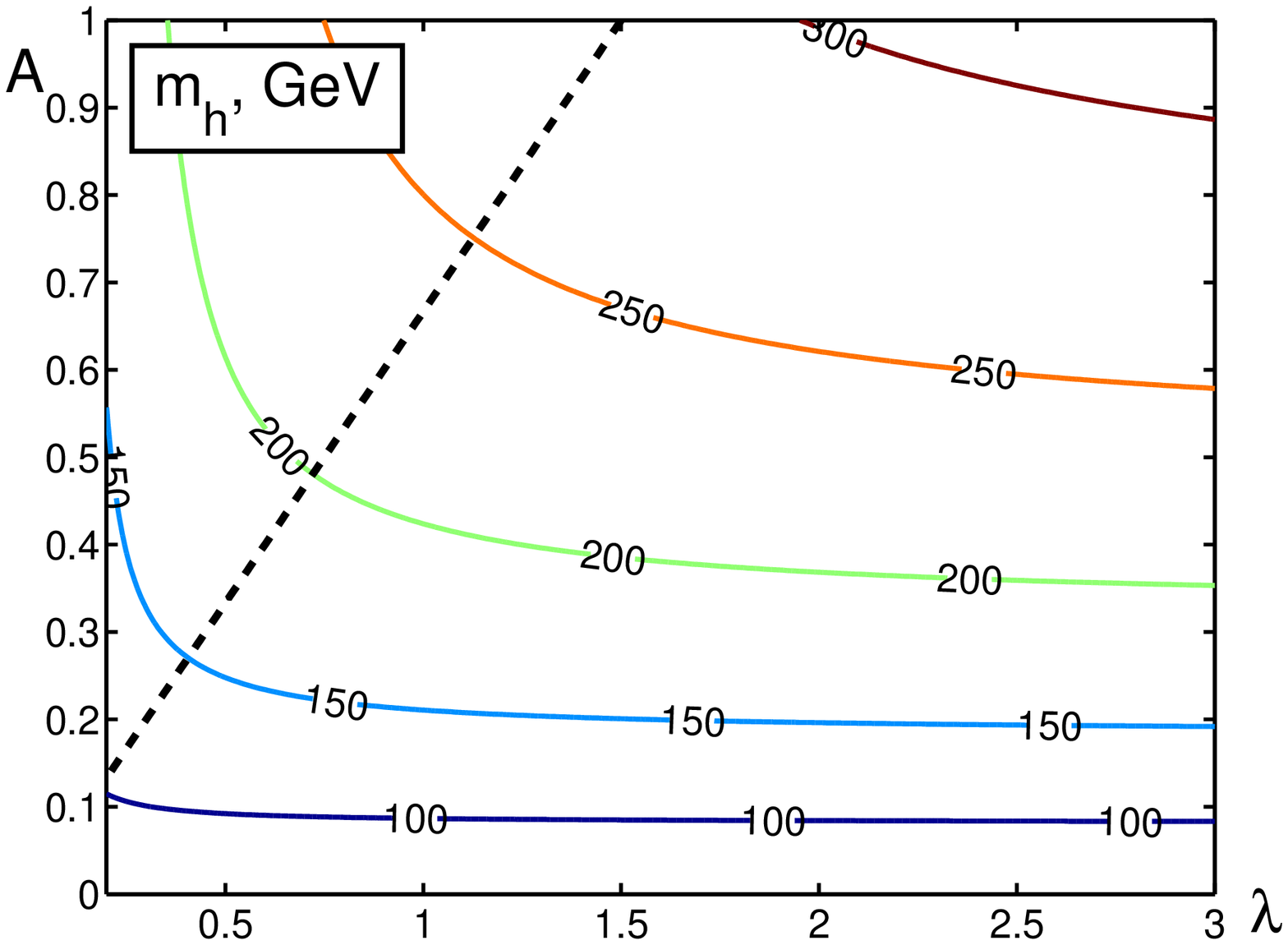} \includegraphics[width=8cm]{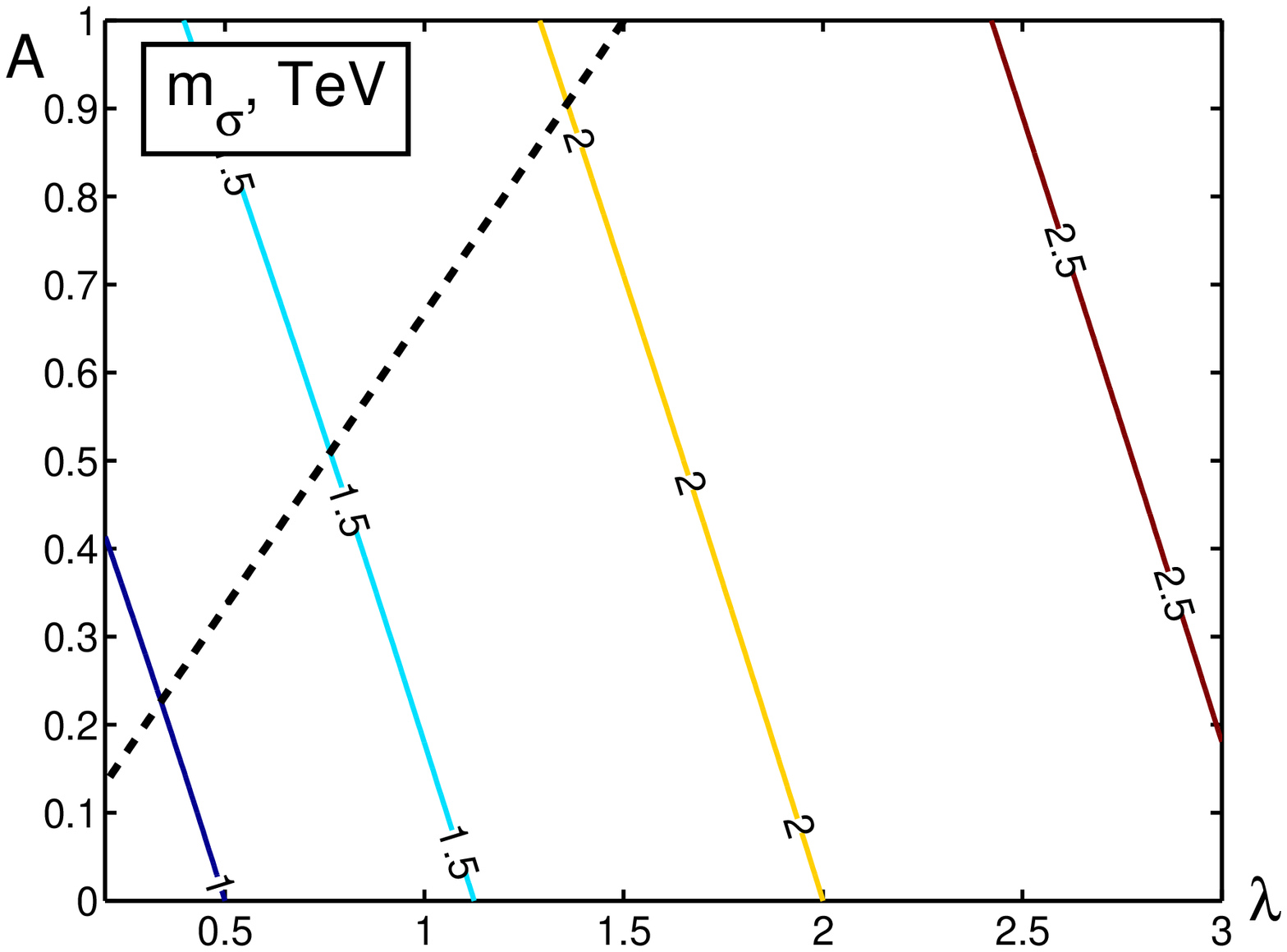}
\end{center}
\caption{{\small ($f=500$ GeV) The masses of the light and heavy scalars in
the perturbative minimal model, Eq.\ (\ref{masses}). The region above the
dashed line corresponds to $z>1/2$ and is relatively disfavored by
Naturalness, see Eq.\ (\ref{ftAA})}.}%
\label{mass-plot}%
\end{figure}

\begin{figure}[h]
\begin{center}
\includegraphics[width=8cm]{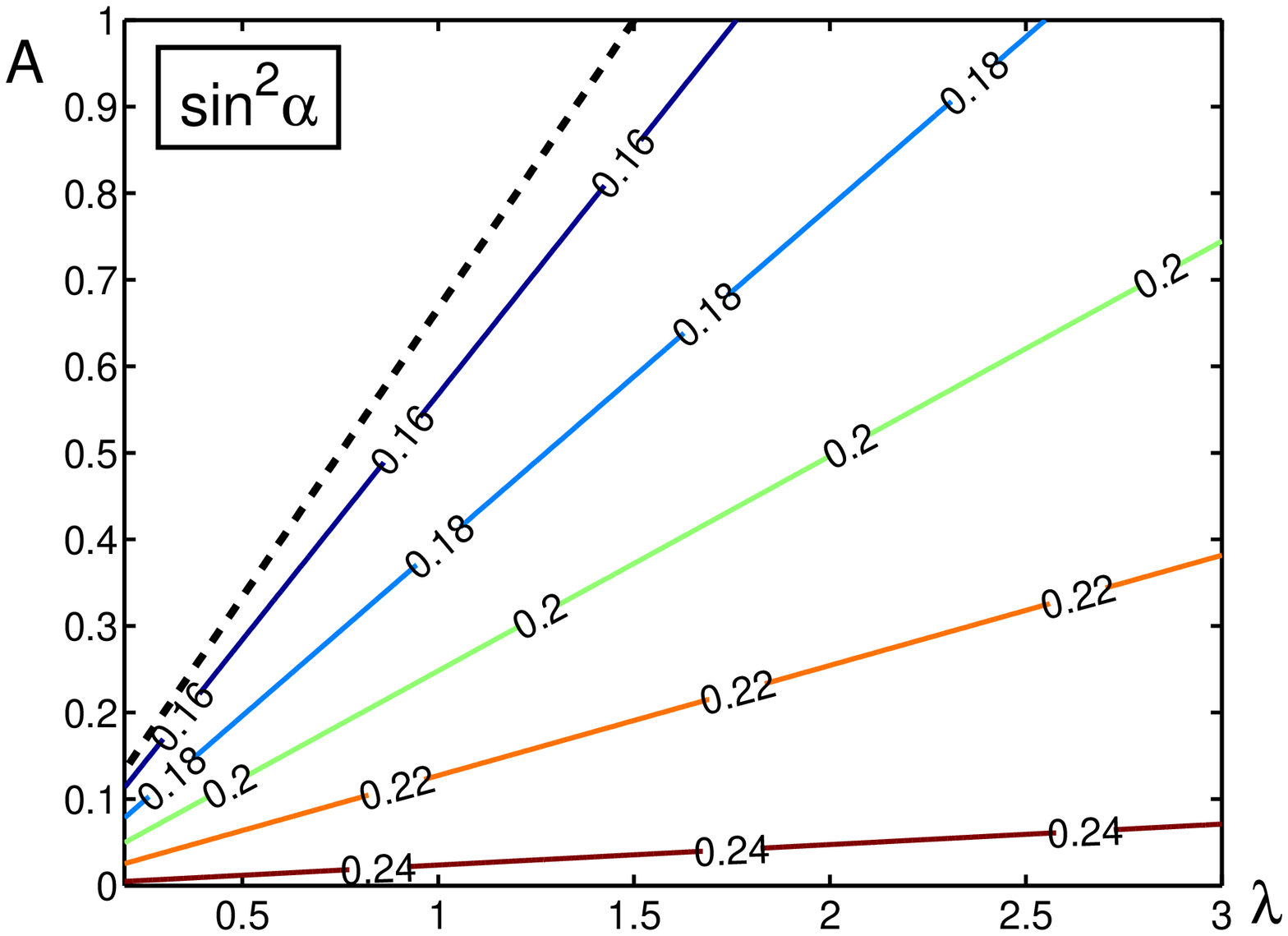} \includegraphics[width=8cm]{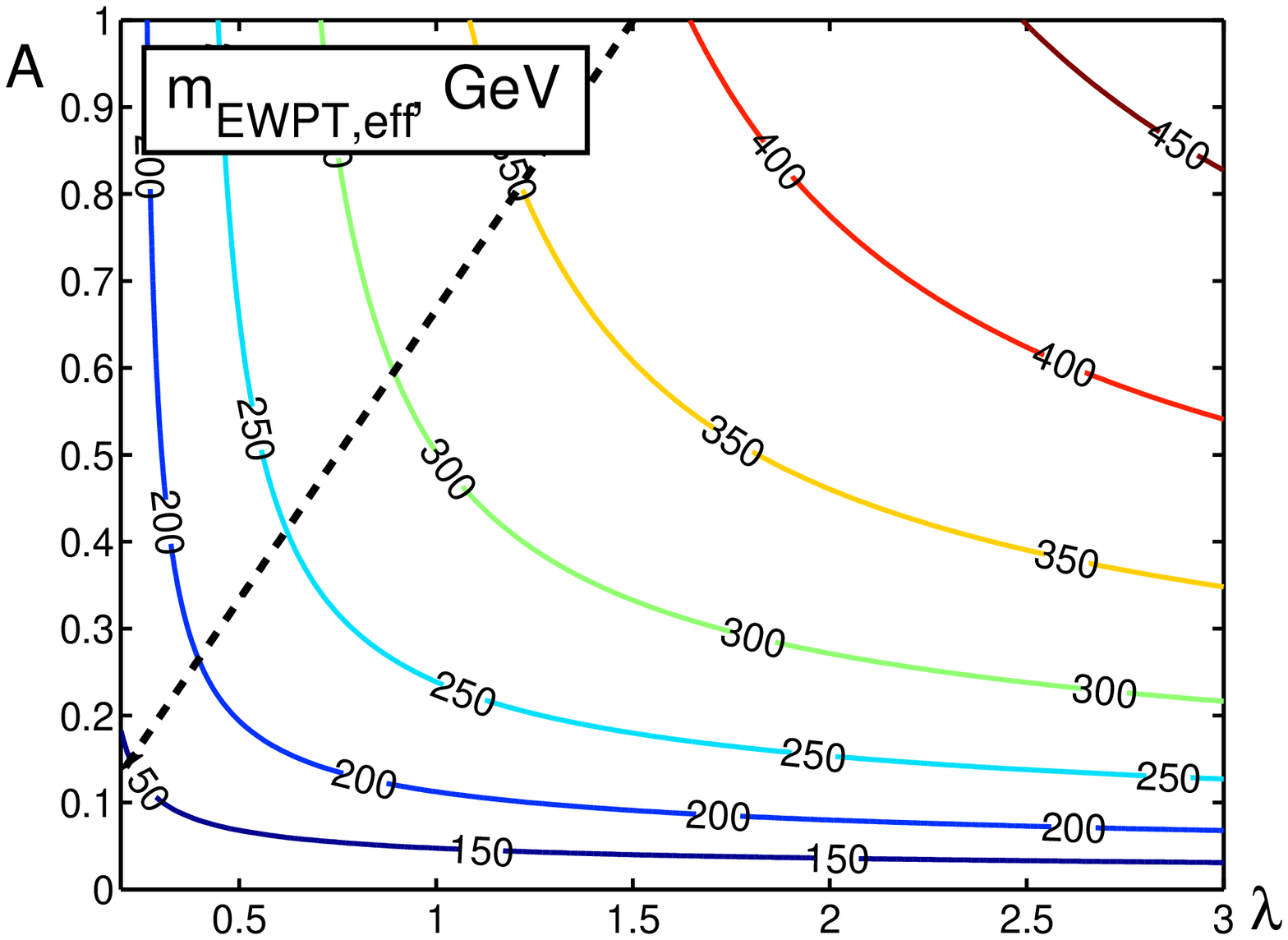}
\end{center}
\caption{{\small ($f=500$ GeV) The mixing angle $\alpha$ between the heavy and
light scalars, see Eq.\ (\ref{scalars}), and the effective EWPT mass as
defined by (\ref{meff}) with $\Lambda$ replaced by $m_{\sigma}$. The region
above the dashed line has the same meaning as in Fig.\ \ref{mass-plot}.}}%
\label{a-m}%
\end{figure}

There are some significant differences with respect to the strongly
interacting case in the EWSB sector, both in the spectrum and in the
couplings. The connection of $v$ to $f$ is now given by
\[
\langle|H|^{2}\rangle\equiv v^{2}=\frac{f^{2}}{2}\left[  1+\frac{A}{2\lambda
}-\left(  \frac{B}{2A}\right)  ^{2}\right]  >0\,,
\]
hence a modified finetuning relation
\begin{equation}
\Delta=\frac{A}{v^{2}}\frac{\partial v^{2}}{\partial A}\simeq\frac{f^{2}%
}{v^{2}}(1+z)\,,\quad~z=\frac{3A}{4\lambda}\,, \label{ftAA}%
\end{equation}
which requires $z$ to be somewhat smaller than unity in order not to worsen
the finetuning. More importantly, the scalar spectrum now contains two scalar
particles below the cutoff
\begin{equation}
h=\cos{\alpha}~\phi_{3}+\sin{\alpha}~\phi_{5},~~~\sigma=-\sin{\alpha}~\phi
_{3}+\cos{\alpha}~\phi_{5}\,, \label{scalars}%
\end{equation}
whose masses, given by
\begin{equation}
m^{2}=4\lambda f^{2}\left(  1+z\pm\left[  1+2z\left(  1-4v^{2}/3f^{2}\right)
+z^{2}\right]  ^{1/2}\right)  \label{masses}%
\end{equation}
are shown in Fig.~\ref{mass-plot} as functions of $A$ and $\lambda$ (for
$f=500$ GeV). Note that $\lambda=3$ makes $m_{\sigma}$ exceed $\Lambda
_{\text{nat}}$. In the same way the mixing angle and the effective value of
$m_{\text{EWPT,eff}}$, to be used as in Section \ref{EWPT} to determine the
corrections to $\hat{S},\hat{T}$, are given in Fig.~\ref{a-m}.

An important consequence of the presence of the $\sigma$ particle below the
cutoff is that the growth of the longitudinal $WW$ cross section as in
(\ref{WW}) is actually cutoff at $\sqrt{s}\simeq m_{\sigma}$, where the
constant behavior sets in. Finally the description of the third-generation
quarks and of their consequences for the EWPT are unchanged after the
following identification between the $f$ parameters of the models at
perturbative and strong coupling:%
\[
f_{\text{strong}}=f_{\text{pert}}\left(  1+\frac{A}{2\lambda}\right)
^{1/2}\,.
\]

We conclude that in a large range of couplings, $\lambda=0.5\div3$, the
perturbative model gives a simple extension of the strong coupling model of
Section \ref{Strong}. Their cutoffs (\ref{Lnat}) and (\ref{cutoff}) have a
different interpretation (in the former case, new physics should cutoff the
quadratic divergence destabilizing the $f$ scale, while in the latter it
should restore unitarity of the longitudinal $WW$ scattering, see
(\ref{unbound})), however numerically they are both close to $3$ TeV. The
finetuning needed to get $v\ll f$ is comparable. Consistency with the EWPT
could be even better than in the strong coupling case, since the contribution
to the S parameter from the cutoff, eq. (\ref{Scutoff}), need not be present
if the physics which cuts off the quadratic divergence of the $\phi$ potential
enters only at loop level\footnote{In strong coupling case, we expect
vectorial resonances at the cutoff, which contribute to the S parameter at
tree level, see examples in Section 6.}. In this case a relatively smaller
$\Delta T\simeq0.1-0.2$ could suffice to restore the consistency of the EWPT
fit. Such a $\Delta T$ could be produced by the extended 3rd generation, eq.
(\ref{deltaTR}), without exceeding experimental constraints in B physics
discussed in Section \ref{Flavor} (although giving effects which could be
observable in the future).

Finally, it could be nontrivial to distinguish the strongly coupled and
perturbative model at the LHC. Even in the most favorable case $m_{\sigma
}\simeq1$ TeV, $(\sin\alpha)^{2}\simeq0.2$ (the lower left corner of the
$\lambda,A$ plane in the plots), the production cross section of the $\sigma$
particle will be only $\sim10$ fb, which makes its observation challenging if
not impossible.

\section{Non-minimal models}

\label{Non-minimal}

\subsection{General picture}

\label{General}As we have already mentioned above, and stress again now, the
model described in Section 2.1 can be considered as a low energy description
of any model in which the EWSB sector has an $SO(5))$ global symmetry partly
gauged by $G_{\text{SM}}$. In all such theories one can introduce an effective
dimensionless fiveplet field $\phi_{\text{eff}},$ $\phi_{\text{eff}}^{2}=1,$
which specifies the alignment angle. Symmetry considerations imply that the
symmetry breaking term in the effective action for the SM gauge fields has to
have the form
\begin{equation}
\mathcal{L}_{\text{EWSB,eff}}=\frac{1}{2}\eta_{\mu\nu}^{\perp}\Pi(p^{2}%
)\phi_{\text{eff}}^{t}A_{\mu}A_{\nu}\phi_{\text{eff}}\,,\quad\eta_{\mu\nu
}^{\perp}=\eta_{\mu\nu}-\frac{p_{\mu}p_{\nu}}{p^{2}}\,, \label{mi}%
\end{equation}
where $A_{\mu}$ is an auxiliary $SO(5)$ gauge field with all of the components
except for the SM gauge fields set to zero:%
\begin{equation}
A_{\mu}=W_{\mu}^{a}T_{L}^{a}+B_{\mu}T_{R}^{3}. \label{aux}%
\end{equation}
The self-energy $\Pi(p^{2})$ depends on the theory under consideration; to
compute it, it is enough to consider the perfect alignment case $\phi
=(0,0,0,0,1).$ This observation is the essence of the so-called method of
matching effectively applied in \cite{Pomarol_spin,Pomarol_fund} in the case
of 5d models. By comparing with Section 2.1, we can identify the effective
sigma-model scale:%
\begin{equation}
f^{2}=\Pi(0)\,, \label{feff}%
\end{equation}
so that the weak scale (or the W mass) is given in terms of the misaligned
VEV
\begin{equation}
\phi_{\text{eff}}=(\varepsilon,0,0,0,\sqrt{1-\varepsilon^{2}}) \label{mis}%
\end{equation}
by the same relation as (\ref{v1}):%
\begin{equation}
v^{2}=\frac{1}{2}\varepsilon^{2}f^{2}. \label{v}%
\end{equation}
For nonzero $\Pi^{\prime}(0),$ the Lagrangian (\ref{mi}) also describes the
kinetic mixing between $W_{3}$ and $B$, i.e.~the $\hat{S}$ parameter. For the
same $\phi_{\text{eff}}$ as in (\ref{v}), we have \cite{Pomarol_spin}%
\begin{equation}
\hat{S}=\frac{g^{2}\varepsilon^{2}}{4}\Pi^{\prime}(0)=\Pi^{\prime}%
(0)\frac{m_{W}^{2}}{f^{2}}\,. \label{sgen}%
\end{equation}

One can imagine that in a general class of models the low energy effective
potential for $\phi_{\text{eff}}$ will have the form of a quadratic polynomial
$f^{4}(A\vec{\phi}_{\text{eff}}^{2}+B\phi_{\text{eff},5})$, analogous to
(\ref{potential}), with parameters $A$ and $B$ functions of more fundamental
parameters of the theory. Assuming that $A$ and $B$ scan their typical ranges,
the finetuning estimate (\ref{ftA}) will apply generally to all models of this
sort. This means that two different models with the same $f$ will likely have
the same level of finetuning, and at this point can be meaningfully compared
with respect to other criteria, such as consistency with the EWPT.

To illustrate the above general points, we consider two concrete examples of
extended models which can be efficiently described by Eqs. (\ref{mi}%
)-(\ref{sgen}). We then compare their `performance' with the minimal models of
Sections \ref{EWSB} and \ref{Perturbative}.

\subsection{Two-site deconstructed model (\textquotedblleft Little Higgs")}

The EWSB sector of this model consists of a real scalar $5\times5$ matrix
field $\Sigma,$ of a scalar fiveplet $\Phi,$ and of an $SO(5)$ gauge field
$X_{\mu}$ which acts on $\Phi$ and, from the right, on $\Sigma$. The $\Sigma$
is assumed to take a VEV, $\langle\Sigma\rangle=f_{0}\boldsymbol{1}.$ There is
a global $SO(5)$ symmetry acting on $\Sigma$ from the left. If $\Phi$ also
takes a VEV, $\Phi^{2}=F^{2}$, this $SO(5)$ global symmetry is broken to
$SO(4).$ We put the SM gauge group inside an $SO(4)$ subgroup of $SO(5)$
acting on $\Sigma$ from the left (without loss of generality, we assume that
this $SO(4)$ acts on the first 4 lines of $\Sigma$). At this point we see that
this model fits the general framework of the previous Section, and thus we
expect that the effective symmetry breaking lagrangian will have the form
(\ref{mi}) with $\phi_{\text{eff}}=\langle\Phi/|\Phi|\rangle$.

We can obtain an equivalent description of the same model by going to a
different gauge in which $\Phi=(0,0,0,0,F).$ In this gauge the Goldstone
degrees of freedom are contained in the matrix field $\Sigma$ subject to the
constraint $\Sigma\Sigma^{t}=f_{0}^{2}\boldsymbol{1}.$ In what follows we
concentrate on the interesting limiting case $F\gg f_{0}$. In this case the
model simplifies, since only the $SO(4)$ subgroup of the $X_{\mu}$ gauge
bosons survives. Thus the model has an extended gauge group $G_{SM}%
\times\lbrack SU(2)_{1}\times SU(2)_{2}],$ with the two new $SU(2)$'s assumed
to have the same coupling $g_{s}$. The EWSB comes from the kinetic Lagrangian
of the $\Sigma$ field:%
\begin{align*}
\mathcal{L}_{\text{kin}}  &  =\frac{1}{2}\text{Tr\thinspace}(D_{\mu}%
\Sigma)(D_{\mu}\Sigma)^{t},\\
\quad D_{\mu}\Sigma &  =\partial_{\mu}\Sigma+i(W_{\mu}^{a}T_{L}^{a}+B_{\mu
}T_{R}^{3})\Sigma-i\Sigma(W_{1\mu}^{a}T_{L}^{a}+W_{2\mu}^{a}T_{R}^{a})
\end{align*}
This model can be obtained as a two-site deconstruction of the 5d model
described below, which explains its name.

Constructions of this type (although perhaps not this precise one) were
extensively discussed in the literature on Little Higgs models \cite{LH}
because they realize the so-called collective symmetry breaking mechanism,
which removes, at the one-loop level, the quadratically divergent contribution
to the Higgs mass parameter from the coupling to the gauge bosons\footnote{The
Little Higgs models in a strict sense of the term also contain a mechanism to
generate the Higgs quartic coupling at tree level. Such a mechanism is absent
in the model under discussion.}. The SM gauge boson loop is canceled by a loop
of new heavy gauge bosons present in the theory.

In principle, as we have seen in Section \ref{Naturalness}, the sensitivity of
the Higgs mass parameter to the gauge boson loop is subdominant to the typical
top-loop contribution, even when the latter is cutoff by extra fermionic
states. In practice this means that naturalness considerations do not require
the presence of states regulating the gauge-boson loop below $2\div3$ TeV,
which weakens the case for their observation at the LHC. Nevertheless, let us
go ahead and analyze the two-site model in some detail. The low energy
effective action for the SM gauge bosons will have the form%
\begin{equation}
\mathcal{L}_{\text{eff}}=-\frac{1}{2}\eta_{\mu\nu}^{\perp}(\frac{p^{2}}%
{g_{0}^{2}}W_{\mu}^{a}W_{\nu}^{a}+\frac{p^{2}}{g_{0}^{\prime2}}B_{\mu}B_{\nu
}+\Pi_{0}(p^{2})\text{Tr}[A_{\mu}A_{\nu}]-\Pi(p^{2})\phi_{\text{eff}}%
^{t}A_{\mu}A_{\nu}\phi_{\text{eff}}). \label{leff-dec}%
\end{equation}
Here the last two terms appear when integrating out the $W_{1,2}$ at tree
level. In agreement with the above discussion, they can be written in an
$SO(5)$-covariant form using the auxiliary field notation (\ref{aux}). The
effective $SU(2)_{L}\times U(1)$ coupling constants $g,g^{\prime}$ at low
energy are given by (see (\ref{mis}))%
\begin{align}
&  \frac{1}{g^{2}}-\frac{1}{g_{0}^{2}}=\frac{1}{g^{\prime2}}-\frac{1}%
{g_{0}^{\prime2}}\,=\delta~,\label{gauge-eff}\\
&  \delta=\Pi_{0}^{\prime}(0)-\Pi^{\prime}(0)\frac{\varepsilon^{2}}%
{4}\nonumber
\end{align}
The symmetric and symmetry-breaking formfactors $\Pi_{0}$ and $\Pi$ are
evaluated by an explicit calculation to be:%
\[
\Pi_{0}=\frac{p^{2}}{g_{s}^{2}}\left(  1-\frac{p^{2}}{g_{s}^{2}f_{0}^{2}%
}\right)  ^{-1},\quad\Pi=2f_{0}^{2}\left(  1-\frac{p^{2}}{g_{s}^{2}f_{0}^{2}%
}\right)  ^{-1}\,.
\]
This gives $\delta=g_{s}^{-2}\left(  1-\varepsilon^{2}/2\right)  $ in
(\ref{gauge-eff}). The couplings $g_{0}$,$g_{0}^{\prime}$, and $g_{s}$ have to
be adjusted so that $g$ and $g^{\prime}$ take their SM values.

Using the general formulas (\ref{feff}) and (\ref{sgen}), we can also compute
the $\hat{S}$ parameter of the model:%
\begin{equation}
\hat{S}=\frac{m_{W}^{2}}{m_{W^{\prime}}^{2}}\simeq1.6\times10^{-3}\left(
\frac{2\,\text{TeV}}{m_{W^{\prime}}}\right)  ^{2} \label{Sdec}%
\end{equation}
where
\[
m_{W^{\prime}}\simeq g_{s}f_{0}%
\]
is the mass of the lightest among the new heavy gauge bosons present in the
theory (these masses can be found as extra zeros of the SM\ gauge bosons
self-energies occurring at $p^{2}>0$). Physically the nonzero $\hat{S}$
parameter appears because of tree-level mixing between the SM gauge bosons and
these new vector states. Going back to the discussion in Section \ref{EWPT},
we see that the EWPT do not allow the heavy gauge bosons below about 2 TeV.

In the above discussion we did not have to make any assumption about the
dynamics which causes $\phi_{\text{eff}}$ to assume a misaligned VEV
(\ref{mis}). Rather generally, one can describe such dynamics in terms of a
soft symmetry breaking potential for $\Sigma$ of the form consistent with the
gauge symmetry of the model:%
\[
Af_{0}^{2}\text{Tr}(\Sigma\boldsymbol{1}_{4\times4}\Sigma^{t}\boldsymbol{1}%
_{4\times4})+Bf_{0}^{3}\Sigma_{55}\,,\quad\boldsymbol{1}_{4\times
4}=\mathrm{diag}(1,1,1,1,0)\,.
\]
In this case, the finetuning of the model can be estimated by (\ref{ftA}) via
the effective sigma-model scale of the model $f$, found from (\ref{feff}) to
be%
\[
f^{2}=2f_{0}^{2}\,\text{.}%
\]

Finally, let us estimate the cutoff of the model. This can be done imagining a
completion into a linear sigma-model with the $SO(5)\times SO(5)$ symmetric
potential%
\[
V=\lambda\text{Tr}(\Sigma\Sigma^{t}-f_{0}^{2}\boldsymbol{1})^{2}%
\]
Demanding that the contribution from self-interaction loops to $f_{0}^{2}$
does not exceed its low-energy value, we get the cutoff ($N=5)$%
\begin{equation}
\Lambda=\frac{4\pi f_{0}}{\sqrt{2N+1}}\simeq3f\,. \label{cutoff1}%
\end{equation}
In the limit of strong $\lambda$ we should use e.g. unitarity bounds to
properly set the cutoff, but we expect that the bounds obtained this way will
be not far from (\ref{cutoff1}), as it happened in the $SO(5)/SO(4)$ case, see
Sections \ref{EWSB} and \ref{Perturbative}. For $f=500$ GeV we get
$\Lambda\simeq1.5$ TeV, which is a rather low value. In particular, the heavy
gauge bosons, in order to be consistent with the EWPT, should have masses
exceeding the cutoff of the theory.

\subsection{5d model}

The model \cite{Pomarol_spin,Pomarol_fund} is defined in flat 5d spacetime
compactified on an interval $0\leq y\leq l.$ It has $SO(5)$ gauge symmetry in
the bulk which is broken to $G_{SM}$ at $y=0$ (the so-called UV brane) and to
$SO(4)$ at $y=l$ (IR brane). The model thus fits the general scheme of Section
\ref{General}. The Lagrangian is%
\begin{align*}
\mathcal{L}  &  =\mathcal{L}_{0}\delta(y)+\mathcal{L}_{5}+\mathcal{L}%
_{l}\delta(y-l)\,,\\
\mathcal{L}_{0}  &  =-\frac{1}{4g_{0}^{2}}(W_{\mu\nu}^{a})^{2}-\frac{1}%
{4g_{0}^{\prime2}}B_{\mu\nu}^{2}\,,\\
\mathcal{L}_{5}  &  =-\frac{M}{4}A_{MN}^{2}\,,\\
\mathcal{L}_{l}  &  =\frac{1}{2}(\partial_{\mu}\Phi-A_{\mu}\Phi)^{2},\quad
\Phi^{2}=F^{2}\,.
\end{align*}
Here $A_{M}$ is an $SO(5)$ gauge field in 5d, which on the UV brane has only
$G_{SM}$ nonzero boundary values (\ref{aux}). The parameter $M$ with dimension
of mass is related to the 5d gauge coupling constant by $M=1/g_{5}^{2}$. The
alignment parameter is $\phi_{\text{eff}}=\Phi/|\Phi|$, the same as in the
previous model. In what follows we consider the limit when $F$ is much bigger
than any other scale in the theory, so that $\mathcal{L}_{l}$ effectively
enforces boundary conditions $A_{\mu}\phi_{\text{eff}}|_{l}=0$.

The original model \cite{Pomarol_spin,Pomarol_fund} was formulated in the AdS
space, with the purpose of resolving the Hierarchy Problem up to the Planck
scale. Since we are interested only in the Little Hierarchy Problem and in the
LHC phenomenology, we here consider a simpler version in flat 5d space. As is
well known \cite{Higgsless}, the curvature of the AdS space can be mimicked by
the kinetic terms for the SM gauge fields on the UV brane contained in
$\mathcal{L}_{0}$.

After integrating out the bulk, the low-energy effective Lagrangian for the SM
gauge bosons takes again the form (\ref{leff-dec}). The formfactors are
however different; they are given by ($p\equiv\sqrt{p^{2}})$%
\[
\Pi_{0}=pM\tan pl,\qquad\Pi=2pM[\tan pl+(\tan pl)^{-1}]
\]
Applying the general formulas (\ref{feff}), (\ref{sgen}), (\ref{gauge-eff}),
we have%
\begin{align}
\frac{1}{g^{2}}  &  =\frac{1}{g_{0}^{2}}+Ml\left(  1-\frac{\varepsilon^{2}}%
{3}\right)  \,,\nonumber\\
f^{2}  &  =\frac{2M}{l}\,,\nonumber\\
\hat{S}  &  =\frac{2}{3}(m_{W}l)^{2}\simeq2\times10^{-3}\left(  \frac
{1.5\text{ TeV}}{l^{-1}}\right)  ^{2} \label{S5d}%
\end{align}
Of interest is the maximal possible value of $M,$ because it controls the
energy cutoff of the 5d theory
\begin{equation}
\Lambda_{\text{NDA}}=\frac{24\pi^{3}M}{N_{c}},\quad N_{c}=5. \label{NDA}%
\end{equation}
Using $l=(1.5$ TeV)$^{-1}$ and $f=500$ GeV (which are the maximal values
affordable without compromising too much with EWPT or Naturalness), we have
$M\simeq80$ GeV$\,,$ corresponding to
\[
\Lambda_{\text{NDA}}\simeq12~\text{TeV}\,,
\]

The heavy vector resonances have masses found from the equation
\[
\frac{p^{2}}{g_{0}^{2}}+\Pi_{0}(p^{2})=0\,.
\]
Since we are in the regime $Ml\ll g_{0}^{-2}$, the first few resonance masses
are well approximated by
\begin{equation}
m_{W^{\prime}}\simeq\frac{\pi}{2l}n\simeq(2.4\text{ TeV)}n\text{, \quad
}n=1,3,5\ldots\label{resonances}%
\end{equation}
It is instructive to rewrite (\ref{S5d}) in terms of the lightest resonance
mass as
\begin{equation}
\hat{S}=\frac{\pi^{2}}{6}\frac{m_{W}^{2}}{m_{W^{\prime}}^{2}} \simeq{1.6}%
\frac{m_{W}^{2}}{m_{W^{\prime}}^{2}}\,. \label{S5d'}%
\end{equation}
The analogous relation in the original AdS model \cite{Pomarol_spin} had a
slightly larger coefficient:
\begin{equation}
\hat{S}_{\mathrm{AdS}}=\frac{27\pi^{2}}{128}\frac{m_{W}^{2}}{m_{W^{\prime}%
}^{2}} \simeq{2.1}\frac{m_{W}^{2}}{m_{W^{\prime}}^{2}}\,. \label{S5dAdS}%
\end{equation}
Comparing (\ref{Sdec}),(\ref{S5d'}),(\ref{S5dAdS}) with (\ref{Scutoff}), we
see that the latter estimate works quite well in all three cases, provided
that we identify the cutoff with the mass of the first resonance.

\subsection{Comparison and appraisal}

Let us conclude this Section with a comparison of the two extended models with
the two minimal models of Sections \ref{EWSB} and \ref{Perturbative}. The
models can only be meaningfully compared at the same level of finetuning,
which in practice means at the same value of the effective sigma-model scale
$f$.

In the two-site model, consistency with the EWPT pushes the heavy gauge bosons
above the cutoff. In such a situation it is hard to see any gain in
introducing the extra gauge bosons in the first place. The calculability of
the theory gets completely lost. In particular, there is no reason to single
out the heavy gauge boson contribution to $\hat{S}$, Eq.~(\ref{Sdec}), among
contributions of other states present at the cutoff.

At the first glance, the situation in the 5d model case is more favourable,
since the ratio of the cutoff and the lightest resonance mass is
$\Lambda_{\text{NDA}}/m_{W^{\prime}}\simeq5$. However, the first resonance
mass is exactly equal, rather than being smaller, to the energy scale
(\ref{unbound}) at which the WW scattering in the effective sigma-model
description exceeds unitarity. In view of this, it appears to us that the
claims about improved calculability in this model have to be substantiated
better than appealing to the NDA estimate (\ref{NDA}), which could be too
optimistic. This could be done, e.g., by computing the one-loop correction to
the tree-level result (\ref{S5d}) for the $\hat{S}$ parameter, and
demonstrating explicitly that this correction is small. Notice also, from
(\ref{resonances}), that the resonances are not equally spaced, so that
already the 3rd resonance mass equals $\Lambda_{\text{NDA}}$.

\section{Conclusions}

The variety of ideas put forward in the context of a \textquotedblleft
composite" picture for the Higgs boson calls for a simple and, at the same
time, effective description of the related phenomenology. In this paper we
attempted to give such a description, and applied it to the potentially
relevant example of a Higgs-top sector from an $SO(5)$ symmetry.

Our starting point is the simple observation that much of the important
phenomenology at relatively low energies should be captured by an approximate
$SO(5)$-invariant Lagrangian obtained by suitably extending the SM Higgs
doublet and the left handed top-bottom doublets: the minimal way is by a real
$5$-plet for the Higgs field and again a $5$-plet of Weyl spinors for the top
doublet, one for each colour, complemented by the three right-handed partners
of the extra components. Other less economic extensions of the third
generation quark-doublet may be considered as well. The $SU(2)\times U(1)$
gauge group of the SM is left untouched.

We believe that this approach is effective in capturing the relevant
phenomenological features of any model based on the same symmetry up to LHC
energies. In particular this makes possible to study in a simple and precise
way the impact of these models on the EWPT as well as on the modified
top-bottom couplings to the gauge bosons. On the basis of this analysis, we
conclude that the minimal $SO(5)/SO(4)$ model with up to $10\%$ finetuning may
be valid up to about $2.5$ TeV, but it has problems in complying with the EWPT
and B-physics constraints. The minimal example that we have analyzed cannot
accommodate the required positive extra contribution to $\delta\rho$ from
fermion loops without introducing at the same time unobserved modifications of
the SM in B-physics. It remains to be seen if this is possible at all in more
extended versions of the 3rd generation quarks. In any event we do not expect
in the spectrum a relatively light $SU(2)$ doublet of hypercharge $7/6$.

All of these considerations do not depend on possible extensions of the gauge
sector. Such extensions, however, can be and have actually been attempted. Our
results apply to them as well, once the symmetry of the EWSB sector and the
description of the top Yukawa coupling are made explicit. Some restrictions of
the parameter space can arise. It is interesting to ask, on the other hand,
which other phenomena may be expected and, especially, if extending the gauge
sector allows to enlarge the domain of the \textquotedblleft minimal" model.
To this end we have considered both a \textquotedblleft little Higgs" two-site
extension and a \textquotedblleft holographic" extension of the $SO(5)/SO(4)$
model. From our results we hardly see any improvement in the \textquotedblleft
little Higgs" case, whereas naive dimensional analysis suggests an extended
range of validity for the \textquotedblleft holographic" model. It is
questionable, however, whether calculability is at the same time maintained.
In our view to assess this issue would require further investigations.

In part for these reasons we have also considered and defended a purely
perturbative version of the \textquotedblleft minimal" model, without any
gauge extension, up to a suitable cutoff, emphasizing the differences with
respect to the strongly coupled case. We noticed that the EWPT consistency
could be better in the perturbative case, if the new physics at the
naturalness cutoff contributes to the S parameter only at loop level. We
believe that the issue of the perturbative versus \textquotedblleft composite"
nature of the Higgs boson should be left as an open (and nontrivial) question
for the experiment to decide.

There are several possible directions for further work along these lines, both
from the point of view of the ``minimal" models and/or of their connections
with ``non-minimal" models. Other symmetries than $SO(5)$ can be considered,
the obvious case being $SU(3)$ (or $SO(6)$). Much of the phenomenological
analyses can be made more precise and explicit. The restrictions arising on
the parameter space of the minimal model from interesting extensions may be
useful to study\footnote{A relevant study has been recently preformed in
\cite{Carena2} for warped holographic extensions, although the modification of
the Higgs contributions to $T$ and $S$ with respect to the SM,
eq.\ (\ref{meff}), has not been taken into account, and the constraints from
$B\bar{B}$ mixing and $b\to s\gamma$ have not been imposed. The issue of
calculability has not been addressed.}. Last but not least, one may try to
address the issue of (partially) UV completing the \textquotedblleft minimal"
model, either by giving a close look at the calculability of existing
proposals or by exploring totally new directions.

\section*{{Acknowledgments}}

{The work of R.B. and V.S.R. was supported by the EU under RTN contract
MRTN-CT-2004-503369. R.B. was also supported in part by MIUR and by a Humbolt
Research Award. We thank Lawrence Hall, Roberto Contino, Adam Falkowski, Gino
Isidori, Guido Martinelli, Hitoshi Murayama, Michele Papucci, Riccardo
Rattazzi, Jos\'{e} Santiago, Marco Serone, Luca Silvestrini, Jesse Thaler, and
Dieter Zeppenfeld for useful discussions. We thank UC Berkeley Physics
department for hospitality while this work was being completed. }


\begin{thebibliography}{99}                                                                                               %


\bibitem {simplified}{ R.~Contino, T.~Kramer, M.~Son and R.~Sundrum,
``Warped/composite phenomenology simplified,'' arXiv:hep-ph/0612180.
}

\bibitem {Rattazzi}{ G.~F.~Giudice, C.~Grojean, A.~Pomarol and R.~Rattazzi,
``The strongly-interacting light Higgs,'' arXiv:hep-ph/0703164.
}

\bibitem {Pomarol_spin}{ K.~Agashe, R.~Contino and A.~Pomarol, ``The minimal
composite Higgs model,'' Nucl.\ Phys.\ B \textbf{719}, 165 (2005)
[arXiv:hep-ph/0412089].
}

\bibitem {Pomarol_fund}{ R.~Contino, L.~Da Rold and A.~Pomarol,
\textquotedblleft Light custodians in natural composite Higgs
models,\textquotedblright\ Phys.\ Rev.\ D \textbf{75}, 055014 (2007)
[arXiv:hep-ph/0612048].
}

\bibitem {Zeppenfeld}Dieter Zeppenfeld, private communication.

\bibitem {elastic}{ B.~W.~Lee, C.~Quigg and H.~B.~Thacker, ``Weak Interactions
At Very High-Energies: The Role Of The Higgs Boson Mass,'' Phys.\ Rev.\ D
\textbf{16}, 1519 (1977); \newline M.~S.~Chanowitz and M.~K.~Gaillard, ``The
Tev Physics Of Strongly Interacting W's And Z's,'' Nucl.\ Phys.\ B
\textbf{261}, 379 (1985).
}

\bibitem {Shat}{ R.~Barbieri, A.~Pomarol, R.~Rattazzi and A.~Strumia,
``Electroweak symmetry breaking after LEP1 and LEP2,'' Nucl.\ Phys.\ B
\textbf{703}, 127 (2004) [arXiv:hep-ph/0405040].
}

\bibitem {IDM}{ R.~Barbieri, L.~J.~Hall and V.~S.~Rychkov, ``Improved
naturalness with a heavy Higgs: An alternative road to LHC physics,''
Phys.\ Rev.\ D \textbf{74}, 015007 (2006) [arXiv:hep-ph/0603188].
}

\bibitem {EWWG}LEP ElectroWeak Working Group, Summer 2006 update,\newline%
\url{http://lepewwg.web.cern.ch/LEPEWWG/plots/summer2006/s06\_stu\_contours.eps
}

\bibitem {Zbb-cust}K.~Agashe, R.~Contino, L.~Da Rold and A.~Pomarol,
\textquotedblleft A custodial symmetry for $Zb\bar{b}$,\textquotedblright%
\ Phys.\ Lett.\ B \textbf{641}, 62 (2006) [arXiv:hep-ph/0605341].


\bibitem {Carena}M.~Carena, E.~Ponton, J.~Santiago and C.~E.~M.~Wagner,
``Light Kaluza-Klein states in Randall-Sundrum models with custodial SU(2),''
Nucl.\ Phys.\ B \textbf{759}, 202 (2006) [arXiv:hep-ph/0607106].


\bibitem {Z-pole}{ ALEPH, DELPHI, L3, OPAL, SLD Collaborations, LEP and SLD
Electroweak Working Groups, and SLD Heavy Flavour Group, ``Precision
electroweak measurements on the Z resonance,'' Phys.\ Rept.\ \textbf{427}, 257
(2006) [arXiv:hep-ex/0509008].
}

\bibitem {Gino}Gino Isidori, private communication.

\bibitem {gambino}M.~Misiak \textit{et al.}, ``Estimate of $B(\bar{B} \to
X_{s} \gamma)$ at $O(\alpha_{s}^{2})$,'' Phys.\ Rev.\ Lett.\ \textbf{98},
022002 (2007) [arXiv:hep-ph/0609232].


\bibitem {UT}Guido Martinelli and Luca Silvestrini, private communication.

\bibitem {Intermediate}E.~Katz, A.~E.~Nelson and D.~G.~E.~Walker, ``The
intermediate Higgs,'' JHEP \textbf{0508}, 074 (2005) [arXiv:hep-ph/0504252].


\bibitem {LH}{ Little Higgs theories. N.~Arkani-Hamed, A.~G.~Cohen and
H.~Georgi, ``Electroweak symmetry breaking from dimensional deconstruction,''
Phys.\ Lett.\ B \textbf{513} (2001) 232 [arXiv:hep-ph/0105239]. For a review,
see M.~Schmaltz and D.~Tucker-Smith, ``Little Higgs review,''
Ann.\ Rev.\ Nucl.\ Part.\ Sci.\ \textbf{55}, 229 (2005)
[arXiv:hep-ph/0502182]; M.~Perelstein, ``Little Higgs models and their
phenomenology,'' Prog.\ Part.\ Nucl.\ Phys.\ \textbf{58}, 247 (2007)
[arXiv:hep-ph/0512128]; R.~Rattazzi, ``Physics Beyond the Standard Model,''
PoS \textbf{HEP2005}, 399 (2006) [arXiv:hep-ph/0607058].
\newline For an efficient introduction with focus on collider phenomenology,
see M.~Perelstein, M.~E.~Peskin and A.~Pierce, ``Top quarks and electroweak
symmetry breaking in little Higgs models,'' Phys.\ Rev.\ D \textbf{69}, 075002
(2004) [arXiv:hep-ph/0310039].
\newline For a concise description of consistency with the EWPT, see
G.~Marandella, C.~Schappacher and A.~Strumia, ``Little-Higgs corrections to
precision data after LEP2,'' Phys.\ Rev.\ D \textbf{72}, 035014 (2005)
[arXiv:hep-ph/0502096].
}

\bibitem {Higgsless}R.~Barbieri, A.~Pomarol and R.~Rattazzi, ``Weakly coupled
Higgsless theories and precision electroweak tests,'' Phys.\ Lett.\ B
\textbf{591}, 141 (2004) [arXiv:hep-ph/0310285].


\bibitem {Carena2}M.~Carena, E.~Ponton, J.~Santiago and C.~E.~M.~Wagner,
``Electroweak constraints on warped models with custodial symmetry,''
arXiv:hep-ph/0701055.

\end{thebibliography}
\end{document}